# IMPROVING COMPUTER ASSISTED SPEECH THERAPY THROUGH SPEECH BASED EMOTION RECOGNITION

Ovidiu A. Schipor

*"Stefan cel Mare" University of Suceava, Universitatii St. 13, RO-720229 Suceava, Romania*
*schipor@eed.usv.ro*

*Abstract: Speech therapy consists in a wide range of services whose aim is to prevent, diagnose and treat different types of speech impairments. One of the most important conditions for obtaining favourable and steady results is the "immersing" of the subject as long as possible into therapeutic context: at home, at school/work, on the street. Since nowadays portable computers tend to become habitual accessories, it seems a good idea to create virtual versions of human SLTs and to integrate them into these devices. However one of the main distinctions between a Speech and Language Therapist (SLT) and a Computer Based Speech Therapy System (CBST) arise from the field of emotion intelligence. The inability of current CBSTs to detect emotional state of human subjects leads to inadequate behavioural responses. Furthermore, this "unresponsive" behaviour is perceived as a lack of empathy and, especially when subjects are children, leads to negative emotional state such as frustration. Thus in this article we propose an original emotions recognition framework - named PhonEM - to be integrated in our previous developed CBST - Logomon. The originality consists in both emotions representation (a fuzzy model) and detection (using only subjects' speech stream). These exceptional restrictions along with the fuzzy representation of emotions lie at the origin of our approach and make our task a difficult and, in the same time, an innovative one. As far as we know, this is the first attempt to combine these techniques in order to improve assisted speech therapy and the obtained results encourage as to further develop our CBST.*

## I. INTRODUCTION

Speech therapy consists in a wide range of services whose aim is to prevent, diagnose and treat different types of speech impairments [1]. Predictions of computers replacing humans have been around for decades. These forecasts were based on both objective and subjective hypothesis. Although the distinction between the computing power and the (emotional) intelligent behaviour has been the subject of many fundamental debates most of the researchers agrees with the necessity of particular models and algorithms instead of general applicable techniques [2].

While the utilisation of computers as SLT's (Speech and Language Therapist) assistants is quite prolific, there are some milestones to be archived in order to obtain independent, autonomous CBSTs. On the top of the "to do list" is emotional intelligence – the ability to perceive and understand emotions and to react accordingly [3]. Emotional computing becomes important especially when subjects are children because they are prone to animism and they treat the CBST as a real playmate [4]. Will the system detect negative affective states such as frustration (e.g. due to much poor results for an exercises type)? What will be the appropriate behaviour a CBST to have when it will understand (hopefully) that child's state of mind?

Since 2005 we have developed Logomon - the first Computer Based Speech Therapy System for Romanian language [5]. It addresses children (from 5 to 9 years old) with dyslalia – the mispronunciation of one or many sounds. This CBST integrates five modules [6], [7]:

- *Monitor Program* (integrated management for all entities related with therapy: children, SLTs, exercises, speech records, results, timetable, and previous scores; it is the connecting link for all below presented modules);
- *3D Articulator Model* (a real-time, 3D, animated model of the pronunciation related organs: lips, cheeks, teeth, and tongue; indicates the correct positions and movements of above mentioned parts of oral cavity in the case of affected sounds);
- *Exercises Player* (over 35 types of interactive exercises for gradual development of phonetic hearing and speech; presents correct samples of speech related with static and animated images in a "game mode" interface);
- *Personal Portable Device* (a PDA that extend therapy in everyday context by playing exercises and offering feed-back; communicates with Monitor Program in order to generate child's evolution history);
- *Fuzzy Expert System* (integrates all information about a subject using fuzzy approach and generates new knowledge related with therapy process such as: number and length of training sessions and number and types of suitable exercises).

After five years of utilisation and base on SLTs', parents' and children's recommendations we have concluded that the most important, to do, qualitative shift is enriching Logomon with emotion recognition skills. That is, generally speaking, the ability to detect emotional state of the subject using different types of information such as: visual (face images), physiological (pulse, skin conductivity, skin temperature) and audio (speech stream). The detection step has to be followed by an appropriate response, a response to be interpreted by the subject as a familiar one (e.g. changing the difficulty level or the exercises type; changing the threshold for rewards, etc.).

In this paper we focus exclusively on detection of emotion states using only speech stream (generally there are short length and poor quality vocal productions) as information channel. In addition, there are more restrictions coming from the Logomon's area of utilisation (the age of the subjects, the particularities generated by the speech impairments, and the assessment context induced by the interaction with the system). These exceptional restrictions along with the fuzzy representation of emotions lie at the origin of our approach and make our task a difficult and, in the same time, an innovative one.

## II. LITERATURE REVIEW

In our previous articles we have synthetized the existing approaches for emotions representation and we have proposed an innovative one, based on fuzzy logic. Therefore, combining *labelling approach* (an emotional state is a specific emotion selected from a predefined list) with *dimensional approach* (an emotional state is a point in a 2 or 3 dimensional space, defined by complementary axes: pleasant/unpleasant, calm/excited, close/open stance) we have obtained the *fuzzy model approach* [8].

Speech is a complex, highly redundant signal that encodes two types of information: the messages for audience in a specific language and paralanguages parameters (e.g. loudness, tempo, intonation, prosody, pitch, etc.). The second category of information is interpreted by the audience and translated into useful but rather subjective knowledge: speaker age, gender, emotional state and so on.

Extensive reviews on emotion recognition from speech reveal significant research concerns [9]: lack of emotions standardisation, language and user dependent systems instead of independent ones, databases with controlled emotions (studio recorded, full blown and/or interpreted by actors), and several unstandardized features to build recognition on. One of the most recent trends refers to attempts to enrich paralanguage information through combining with semantic features (the meaning of the message) [10].

There are three possibilities to obtain speech corpora for emotion recognition system training: simulation (usually using actors), induction (common people are transposed in a specific emotional state), and in a natural, spontaneous way. Each approach has its own pros and cons and is closely related to the field of use. That is why, so called "generally purpose" databases have to be carefully chosen. However, there are several more parameters to focus on such as: speaker (number, gender, age), utterance (number, length, standardisation) and environment (noise level, real life/artificial) related parameters.

The number of recognized emotional states ranges from 2 (negative/positive [11], happiness/neutral [12]) to 7 (anger/boredom/disgust/fear/joy/neutral/sad [13]) and the most three studied emotions are anger, neutral, and happiness. Besides determining specific emotions, several studies have focused on determining the level of emotions. For example, the question "Is the subject stressed?" has been completed with "What is the subject's stress intensity?" [14] [15]. In another researches, the emotions are replaced by several unstandardized "states of mind" such as approval, attention and prohibition [16].

An extensive and standardized comparison between the emotion recognition systems' performances seems to be an overwhelming task due to the large number of variables a specific system is built on (speech database, features, classifiers types). The discussion is even livelier because the benchmark itself is a subjective one. Indeed, subjective listening tests reveal an average recognition performance of about 80% with a 90% peak for studio recorded, acted, and clearly expressed emotions. We cannot expect a computer to outperform the common people's emotion recognition skills. The most important variable seems to be the number of emotions since the recognition rate varies from about 50% (7 emotions) to around 85% (2 emotions).

Although we did not found examples of emotional capable CBSTs, several related field have been studied from this perspective [17][18][19]: E-tutor applications, medical diagnosis tools, and conversation with robotic pets. All the above mentioned authors highlight the important role the emotional intelligent behaviour could play in order to enrich human-computer interaction. In addition, an adapted emotional response (i.e. emotions synthetized by the system in the target speech) leads to a more human-like perceived conversation.

### III. PhonEM ARCHITECTURE AND INTEGRATION

Our speech database consists in about 1500 utterances, human recognized from both pronunciation and emotional point of view. There are some particularities that give our attempt an important role:
- the recording of utterances was done in a common environment;
- subjects: 10 children aged between 5 and 9 years old;
- the average speech fragments length is relatively short (around one second);
- five gradual ordered basic emotions was used for an emotional state representation (negative: nervousness and tenseness, neutral, positive: contentment, happiness);
- the emotions was neither simulated nor inducted, but natural expressed.

The utterances come from three distinct modules of Logomon CBST: Monitor Program (speech assessment), 3D models, and Personal Portable Device (Exercises). There are either already recognized or new utterances corresponding to training and recognition scenarios. PhonEM module performs a features extraction task followed by a HMM training or classification.

The distribution of the emotional state is a biased one (positive affective state – 66%, neutral – 22%, and negative affective state 12%) and, due ethical reason, no external intervention was conducted to correct this. A more detailed analysis shows a specific distribution pattern for different therapeutic stages (Table 1).

TABLE I. The Distribution of the Emotional States.

| Therapeutic steps | negative | neutral | positive |
|---|---|---|---|
| evaluation exercises | 11% | 26% | 63% |
| phonematic hearing developement | 14% | 17% | 70% |
| pronunciation using 3D model | 12% | 20% | 68% |

The majority of training phonemes comes from the speech assessment activity while, on the other hand, the exercises provide utterances that need to be recognized. The speech database is used for storing correspondences between utterances and fuzzy numbers with the mention that more than one numbers can be associated with the same speech fragment. This special non-bijective relation can appear when different human experts provide strong different fuzzy numbers. In this case, we prefer to store all possible emotional states instead trying to obtain a mediated value.

## IV. CONCLUSIONS

Taking into consideration the particularities of a CBST, the detection of the affective state using speech stream seems to be a natural, reliable and non-intrusive approach. Many recent research has been conducted in this field and several strategies has been employed for emotion generation (acted, induces, naturally expressed), feature extraction (spectral and/or prosodic), and classification (linear/non-linear).

We have presented our speech corpora together with an innovative fuzzy model for the emotional state representation. Specific relations have been established between the PhonEM framework and the other modules of the Logomon. This integration takes into account various constraints such a system entails: the subject's age, the quality of the speech (from both noise and pronunciation point of view), and the particularities of utterances (short length, naturally expressed).

There are many future directions to work on. First of all, our approach needs to be validated compering artificial with human classification. Furthermore, a transition from speaker dependent to speaker independent recognition has to be made. A system limited to a user dependent approach will have a poor emotion recognition performance even in the case of the same speaker, while, at this age, the voice particularities and the quality of pronunciation greatly varies. Although these challenges seem overwhelming, the already achieved milestones encourage us to continue our work.

### Reference Text and Citations